# Ideal Bandgap in a 2D Ruddlesden-Popper Perovskite Chalcogenide for Single-junction Solar Cells

Shanyuan Niu[†], Debarghya Sarkar[‡], Kristopher Williams[§], Yucheng Zhou[†], Yuwei Li[#], Elisabeth Bianco[£], Huaixun Huyan[†], Stephen B. Cronin[‡], Michael E. McConney[Π], Ralf Haiges[&], R. Jaramillo[€], David J. Singh[#], William A. Tisdale[§], Rehan Kapadia[‡], Jayakanth Ravichandran*[,†,‡]

[†]Mork Family Department of Chemical Engineering and Materials Science, University of Southern California, Los Angeles, CA 90089, USA

[‡]Ming Hsieh Department of Electrical Engineering, University of Southern California, Los Angeles, CA 90089, USA

[§]Department of Chemical Engineering, Massachusetts Institute of Technology, Cambridge, MA 02139, USA

[#]Department of Physics and Astronomy, University of Missouri, Columbia, MO 65211, USA

[£]Department of Chemistry, Rice University, Houston, TX 77005, USA

[Π]Materials and Manufacturing Directorate, Air Force Research Laboratory, Wright-Patterson AFB, OH 45433, USA

[&]Loker Hydrocarbon Research Institute and Department of Chemistry, University of Southern California, Los Angeles, CA 90089, USA

[€]Department of Materials Science and Engineering, Massachusetts Institute of Technology, Cambridge, MA 02139, USA

**ABSTRACT:** Transition metal perovskite chalcogenides (TMPCs) are explored as stable, environmentally friendly semiconductors for solar energy conversion. They can be viewed as the inorganic alternatives to hybrid halide perovskites, and chalcogenide counterparts of perovskite oxides with desirable optoelectronic properties in the visible – infrared part of the electromagnetic spectrum. Past theoretical studies have predicted large absorption coefficient, desirable defect characteristics, and bulk photovoltaic effect in TMPCs. Despite recent progresses in polycrystalline synthesis and measurements of their optical properties, it is necessary to grow these materials in high crystalline quality to develop a fundamental understanding of their optical properties and evaluate their suitability for photovoltaic application. Here, we report the growth of single crystals of a two-dimensional (2D) perovskite chalcogenide, $Ba_3Zr_2S_7$, with a natural superlattice-like structure of alternating double-layer perovskite blocks and single-layer rock salt structure. The material demonstrated a bright photoluminescence peak at 1.28 eV with a large external luminescence efficiency of up to 0.15%. We performed time-resolved photoluminescence spectroscopy on these crystals and obtained an effective recombination time of ~65 ns. These results clearly show that 2D Ruddlesden-Popper phases of perovskite chalcogenides are promising materials to achieve single-junction solar cells.

Organic-inorganic halide perovskites have been widely studied as semiconductors with extraordinary optoelectronic properties.[1-3] Within a few years of development, solar cells based on halide perovskites have reached power conversion efficiency up to 22.7%,[4-7] representing a major achievement in the rapid development of innovative functional materials. High efficiency hybrid halide perovskites are often composed of rare, and/or toxic elements such as lead and despite the experimental progress, their long term stability compared to other materials remains an open issue.[8] Theoretical studies have proposed transition Metal Perovskite Chalcogenides (TMPCs) as a new class of semiconductors with desirable properties for solar energy conversion.[9-13] Although such perovskite chalcogenides have been synthesized,[14-17] an understanding of their optical properties and suitability for photovoltaics remains largely unexplored. Similarly, if their stability in electrochemical environments could be verified, they can supplant the widely used oxide perovskites, which require irradiation with ultra-violet light for solar water splitting applications.[18]

Recently, several research groups reported optical properties of perovskite chalcogenide phases such as $CaZrS_3$, $BaZrS_3$, $\alpha$-$SrZrS_3$, $\beta$-$SrZrS_3$.[19-22] Although preliminary luminescence studies showed the promise of such materials for photovoltaics,[21] the demonstration of bandgap tunability down to the solar optimal single-junction value remains an outstanding challenge. Several approaches such as alloying,[13,19] and the exploration of quaternary chalcogenide[23] and other non-transition metal-based perovskite chalcogenides[24] have been proposed to overcome this hurdle. While efforts to achieve optimal bandgap for single-junction solar cells and the demonstration of solar cells are underway, other approaches to add new functionalities to this family of materials are being considered. For example, the introduction of ferroelectricity or any static polar order in a semiconductor can lead to interest-ing physical effects such as shift currents.[25] Several theoretical studies explored the possibility of achieving static polarization in the Ruddlesden-Popper type 2D layered TMPCs, to demonstrate bulk photovoltaic effect.[11,26] Thus, studies of optical properties of 2D layered TMPCs are of broad interest beyond their applications as photovoltaic absorbers.

Ruddlesden-Popper phases are 2D homologues series of the perovskite structure. Such layered structures can host interesting octahedral rotations and distortions that can lead to non-centrosymmetric structure, which is a perquisite for both polar nature or ferroelectric properties. 2D perovskites chalcogenides are formed by alternating a set number ($n$) of perovskite layers with the chemical formula $ABX_3$ and a rock salt layer AX. Such a 2D perovskite has a general formula of $A_{n+1}B_nX_{3n+1}$ for the case of the same cations in perovskite and rock salt layer. $Ba_3Zr_2S_7$ is an $n=2$ Ruddlesden-Popper phase of the perovskite sulfide $BaZrS_3$. Two adjacent perovskite layers with corner-sharing $ZrS_6$ octahedra are intercalated by one BaS layer, as shown in **Figure 1(a)**. Motivated by the theoretical studies discussed above, we performed in-depth first principles calculations and experimental studies on the single crystals of $Ba_3Zr_2S_7$ to understand their opto-electronic properties. Surprisingly, our studies showed that $Ba_3Zr_2S_7$ possessed a bandgap of 1.28 eV, which is promising for building single-junction solar cells.

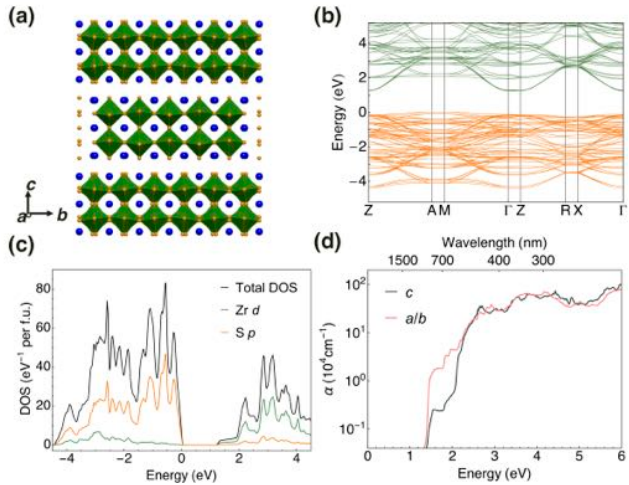

**Figure 1.** (a) Schematic crystal structure of $Ba_3Zr_2S_7$. The blue, yellow and green spheres represent Ba, S and Zr atoms respectively. The $ZrS_6$ octahedra are highlighted in green. (b) Band structure of $Ba_3Zr_2S_7$ calculated with mBJ potential. (c) Total density of states plot and contribution of Zr $4d$ and S $3p$ orbitals. (d) Calculated absorption coefficients along different directions.

In this article, we report the demonstration of the optimal bandgap for single-junction solar cell in the single crystals of a Ruddlesden-Popper phase TMPC, $Ba_3Zr_2S_7$. The single crystals were grown by salt flux method. Structural and chemical characterizations including X-ray diffraction (XRD), Raman spectroscopy, transmission electron microscopy (TEM), and energy dispersive analytical X-ray spectroscopy (EDS) established the crystalline quality. We also performed static, quantitative, and time-resolved photoluminescence studies to evaluate the suitability of $Ba_3Zr_2S_7$ as absorber in a single-junction solar cell.

We performed density functional calculations for $Ba_3Zr_2S_7$ done using the general potential linearized augmented planewave (LAPW) method[27] as implemented in the WIEN2K code.[28] The electronic structure and optical properties were calculated with the modified Becke Johnson (mBJ) potential, which generally improves the accuracy of band gap prediction for semiconductors.[29] The contribution of spin-orbit coupling was included to improve the accuracy of the calculations. The calculated band structure, density of states (DOS), and absorption coefficients are shown in **Figure 1(b), 1(c), 1(d)**, respectively. The calculated electronic structure showed an indirect bandgap of 1.25 eV with a valence band maximum at M point, and a direct gap of 1.35 eV at $\Gamma$ point. Similar to the other early transition metal perovskite oxides and chalcogenides, the conduction and valence band of $Ba_3Zr_2S_7$ are primarily composed of Zr $d$ orbitals and S $p$ orbitals, respectively. The high DOS is also manifested in the flat valence band maximum and conduction band bottom in the band structure. These lead to a sharp absorption onset and large absorption coefficients greater than $10^4$ cm$^{-1}$ near the band edge. It is worth noting that in addition to the structure with octahedral rotation shown here, we also performed calculations for another higher temperature phase with no octahedral rotation, the corresponding indirect and direct band gap were 0.96 and 1.26 eV respectively (more details are available in Supporting Information).

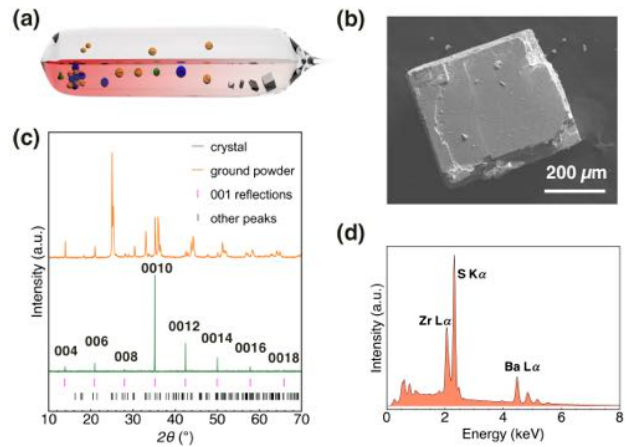

**Figure 2.** (a) Schematic of the salt flux crystal growth. (c) SEM image of a $Ba_3Zr_2S_7$ crystal. (b) XRD of the crystal and the ground powder. The 00$l$ reflections and other reflections are indicated with cyan and black bars. (d) EDS spectrum of a $Ba_3Zr_2S_7$ crystal showing ratios around 1.5:1:3.1. The most intense peaks from Ba, Zr and S are labeled.

The single crystals were grown in sealed quartz ampoules with $BaCl_2$ flux, using a synthetic procedure similar to ones reported earlier.[30] A schematic of the crystal growth is shown in **Figure 2(a)**. The use of appropriate high temperature salt-based solution lowers the crystal growth temperature and enables the single crystal growth from this melt solution in a few days. The predominant morphology of the obtained crystals was cubes and cuboids with well-defined surfaces that presumably correspond to crystal facets as shown in the SEM image (**Figure 2(b)**). X-ray diffraction (XRD) scans of a cuboid crystal and the ground powders (orange) are overlaid in **Figure 2(c)** with expected Bragg reflections. The ground powder indicated phase pure $Ba_3Zr_2S_7$ within the limits of instrument detection, and the out-of-plane XRD on the crystal showed only one set of sharp 00$l$ reflections. The most intense peak has a full-width-at-half-maximum (FWHM) of less than 0.04°. This confirmed that the crystal facets with layered-like features are along the (001) plane. We performed single-crystal XRD studies and found that the crystals adopted the $P4_2/mnm$ space group. The deduced lattice constants were $a = 7.079(2)$ Å, $b = 7.079(2)$ Å, and $c = 25.437(5)$ Å, respectively (note that the in-plane crystallographic axes $a$ and $b$ denoted here are 45° rotated from the pseudo-cubic notation shown in Figure 1(a), complete structural parameters are available in the Supporting Information), which agree well with previous structural studies.[30] EDS with varying locations and magnifications on the crystals showed only expected elements and a consistent ratio of Ba:Zr:S around 1.5:1:3.1. A representative EDS spectrum on the crystal surface is shown in **Figure 2(d)**.

To verify the superlattice-like structure in $Ba_3Zr_2S_7$, we performed scanning transmission electron microscopy (STEM) studies on the crystals. A perspective schematic view of this layered structure is shown in **Figure 3(a)**. The STEM image of a $Ba_3Zr_2S_7$ crystal and corresponding selected area electron diffraction (SAED) pattern are shown in **Figure 3(b)**. As expected, we observed the double-layer perovskite slabs sandwiching the extra rock-salt atomic layer offset by half a unit cell along the face diagonal of the in-plane square lattice. Much denser diffraction spots along $c$-axis correspond to the much larger lattice constant. The highly symmetric SAED pattern also establishes high quality of the single crystal.

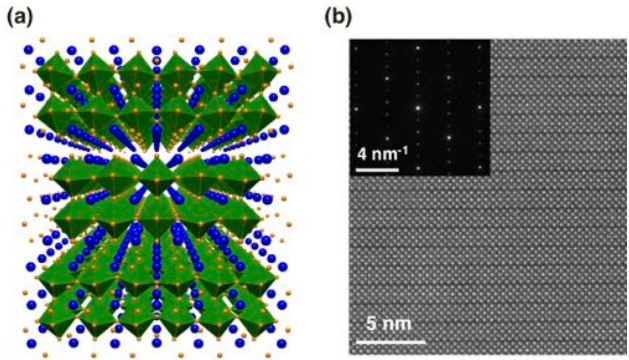

**Figure 3.** (a) A perspective view of the natural superlattice nanostructure in $Ba_3Zr_2S_7$. The blue, yellow and green spheres represent Ba, S and Zr atoms respectively. The $ZrS_6$ octahedra are highlighted in green. (b) STEM image of a $Ba_3Zr_2S_7$ crystal viewed along a-axis, the inset is the corresponding SAED pattern.

The static photoluminescence (PL) measurement on the crystals at room temperature with 785 nm excitation showed a clear and intense emission peak at 1.28 eV. This value is in agreement with the theoretical prediction of ~ 1.35 eV and is presumably due to the direct gap band-band transition. In addition to the optimal band gap, it is necessary to evaluate the material's potential for achieving power conversion efficiency close to the Shockley–Queisser (SQ) limit. The external luminescence efficiency ($\eta_{ext}$), defined as the ratio of output to incident photon numbers, is one of relevant material parameters for this evaluation.[31] Inefficient external luminescence at open circuit is an indicator of non-radiative recombination and optical losses, and $\eta_{ext}$ is thus a thermodynamic measure of the available open-circuit voltage in photovoltaic devices.[32,33] A comparison of the radiative emission of $Ba_3Zr_2S_7$ with reference single crystalline InP and GaAs wafers under the same illumination and measurement conditions is shown in **Figure 4(a)**. Despite the non-ideal surface quality of the flux-grown crystals, integrated emission intensity from $Ba_3Zr_2S_7$ is within one order of magnitude compared to the atomically smooth state-of-art III-V wafers. Measurement of $\eta_{ext}$ was carried out using a 785 nm laser excitation with known incident photon flux, and the output photon flux was measured in a calibrated PL system. The open circuit voltage ($V_{OC}$) under illumination is extracted via $\eta_{ext}$ using method described elsewhere[33]. The measured emitted photon flux at different incident photon flux and extracted $V_{OC}$ under different illumination powers are shown in **Figure 4(b)**. The external luminescence efficiency remains 0.1% ~ 0.15% for incident power from around $10^3$ W m$^{-2}$ to $10^6$ W m$^{-2}$, but quickly drops to below 0.1% as power further increases. This can be understood by considering the relative rates of Shockley-Read-Hall (SRH), radiative, and Auger recombination pathways, which are proportional to the first order, second order and third order of carrier density. External luminescence efficiency will decrease when non-radiative Auger recombination dominates as large excess carrier density is generated at high incident photon flux.

We also performed time-resolved photoluminescence (TRPL) measurements at room temperature to study the non-equilibrium carrier dynamics in $Ba_3Zr_2S_7$. PL was excited with a pulsed diode laser at a wavelength ($\lambda_{PUMP}$) of 405 nm, and the PL transient was measured as a function of time and energy in the range $\lambda_{DETECT}$ = 900–1050 nm using spectrally-resolved, time-correlated single photon counting (TCSPC). In **Figure 4(c)** we show the spectral transient PL decay. In **Figure 4(d)** we show the data integrated over the range $\lambda_{DETECT}$ = 910-1010 nm (peak position ± 50 nm) to select band-to-band radiative recombination, in order to estimate the free carrier recombination lifetime. Using the known pump parameters and the calculated absorption coefficient of $Ba_3Zr_2S_7$, we estimate the peak carrier concentration to be above $10^{19}$ cm$^{-3}$ within 30 nm of the illuminated top surface of the crystal. Therefore, Auger, surface, and bulk SRH are all likely relevant recombination mechanisms. Many material parameters of $Ba_3Zr_2S_7$ are as-yet unmeasured, including the electron and hole diffusivities, equilibrium carrier concentration and type, and the Auger coefficients. Absent such parameters, it is not possible to determine specific recombination rates from the data in Figure 5(b). An empirical, double-exponential convolved with the laser pulse width (0.59 ± 0.04 ns FWHM) fit to the data yields time constants $\tau_1$ = 4.5 ± 0.2 ns and $\tau_2$ = 65 ± 2 ns. The longer time constant is an effective parameter that represents both bulk and surface recombination, and is typically shorter than the bulk SRH recombination lifetime.[34] Therefore, the data suggests that the room-temperature SRH lifetime in our $Ba_3Zr_2S_7$ crystals is well over 60 ns. This is a promising result, comparable already to state-of-art CdTe, CIGS, and halide perovskite materials that support solar cells with one-sun power conversion efficiency over 20%.[35]

In conclusion, we report a Ruddlesden-Popper perovskite chalcogenide, $Ba_3Zr_2S_7$, with an optimal bandgap for single-junction photovoltaic devices and large absorption coefficient near the band edge. Flux grown single crystals showed a bright photoluminescence at 1.28 eV, with an external luminescence efficiency up to 0.15%. The effective minority carrier recombination time determined via TRPL measurement was 65 ns. The combination of optimal band gap, short absorption length, high external luminescence efficiency, and long carrier lifetime make this inorganic layered perovskite chalcogenide promising candidate for photovoltaic applications. These findings establish the growing potential of transition metal perovskite chalcogenides as semiconductors for broad optoelectronic applications.

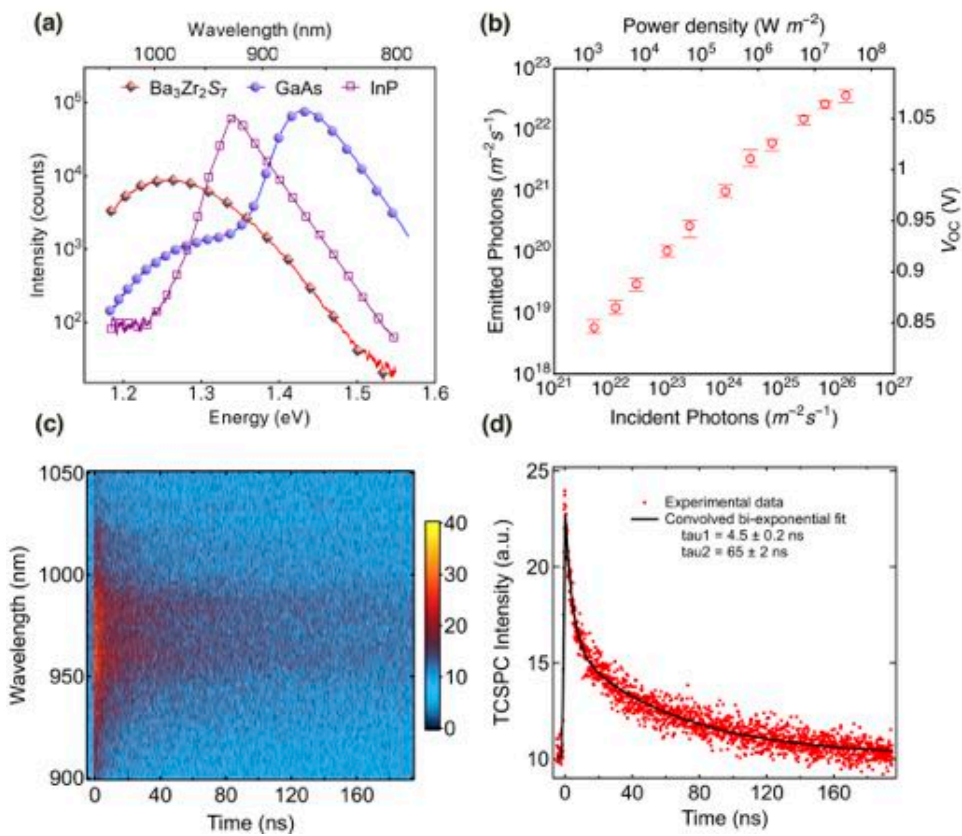

**Figure 4.** (a) PL intensity comparison of a $Ba_3Zr_2S_7$ crystal, a InP wafer and a GaAs wafer under the same measurement conditions. (b) Quantitative emission and corresponding $V_{OC}$ at different incident power density. The error bar is included as multiple sets of data were obtained on several $Ba_3Zr_2S_7$ crystal pieces. (c) Spectral- and time-resolved emission map. The TCSPC intensity is indicated with color. (d) Intensity decay profile of the emission peak as a function of time. The solid line is the bi-exponential fit convolved with the pump profile. A fast and slow decaying time constant of 4.5 ns and 65 ns are extracted.

## ASSOCIATED CONTENT

### Supporting Information

The Supporting Information is available free of charge on the ACS Publications website.

More experimental details of DFT calculations, crystal growth, powder XRD, electron microscopy, single crystal XRD, extraction of $V_{OC}$, and time-resolved PL are available in the supplementary information (PDF).
Crystallographic information file (cif).

## AUTHOR INFORMATION

### Corresponding Author

* Jayakanth Ravichandran: jayakanr@usc.edu

## ACKNOWLEDGMENT

J.R. and S.N. acknowledge USC Viterbi School of Engineering Startup Funds and support from the Air Force Office of Scientific Research under award number FA9550-16-1-0335. S.N. acknowledges Link Foundation Energy Fellowship. R.K acknowledges support from National Science Foundation under award number 1610604. D.S. acknowledges the USC Annenberg Graduate Fellowship. K.W. and W.A.T. were supported by the U.S. Department of Energy, Office of Science, Office of Basic Energy Sciences, under award number DE-SC0010538. Work at the University of Missouri was supported by the Department of Energy through the MAGICS center, Award DE-SC0014607. E.B. acknowledges support from the Electron Microscopy Center in the Shared Equipment Authority at Rice University and the National Science Foundation Graduate Research Fellowship under Grant No. (DGE-1450681). M.E.M. and E.B. acknowledges support by the Air Force Office of Scientific Research under award number FA9550-15RXCOR198. The authors gratefully acknowledge the use of Center for Electron Microscopy and Microanalysis at University of Southern California for materials characterization.

Insert Table of Contents artwork here

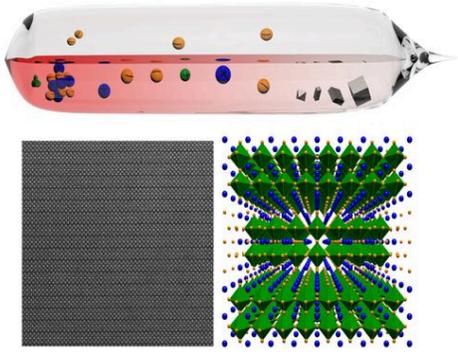